\documentclass[11pt,a4paper]{article}
\usepackage[left=15mm, top=10mm, right=15mm, bottom=20mm, nohead=true, nofoot=true]{geometry}

\usepackage[utf8]{inputenc}
\usepackage{authblk}
\usepackage{indentfirst}
\usepackage{misccorr}
\usepackage{graphicx}
\usepackage{amsmath}
\usepackage{amssymb}
\usepackage{multicol}
\usepackage{bm}
\usepackage{placeins}
\usepackage{longtable}
\usepackage{pdflscape}
\usepackage{epstopdf}
\usepackage{multirow}
\usepackage{adjustbox}
\usepackage{amsfonts}
\usepackage{ifthen}
\usepackage{fancyhdr}
\usepackage{setspace}
\usepackage{xcolor}
\usepackage[round,authoryear,comma,sort&compress,numbers]{natbib}
\usepackage{subcaption} % Add this to your preamble
\usepackage[toc,title,page]{appendix}
\usepackage[hidelinks]{hyperref}
\usepackage{url}

\hypersetup{
    colorlinks=true,
    linkcolor=green,
    citecolor=blue,
    filecolor=magenta,
    urlcolor=cyan,
    pdftitle={Sharelatex Example},
    pdfpagemode=FullScreen,
}

\fancypagestyle{plain}{\chead{\texttt{\textcolor{brown}{Communications of BAO, Vol.~72, Issue~II, 2025, pp.~}}}}
\title{\textbf{Metallicity Effects on Machine Learning Classification of Dusty Stellar Sources in the Magellanic Clouds}}

\author[1]{Sepideh Ghaziasgar \thanks{sepideh.ghaziasgar@ipm.ir, Corresponding author}}
\author[1]{Mahdi Abdollahi 
\thanks{m.abdollahi@ipm.ir}}
\author[1]{Atefeh Javadi 
\thanks{atefeh@ipm.ir}}
\author[2]{Jacco Th. van Loon \thanks{j.t.van.loon@keele.ac.uk}}
\author[3]{Iain McDonald \thanks{Iain.Mcdonald-2@manchester.ac.uk}}
\author[2]{Joana Oliveira \thanks{j.oliveira@keele.ac.uk}}
\author[1,4]{Habib G. Khosroshahi \thanks{habib@ipm.ir}}
\affil[1]{\scriptsize School of Astronomy, Institute for Research in Fundamental Sciences (IPM), P.O. Box 19568-36613, Tehran, Iran}
\affil[2]{\scriptsize Lennard-Jones Laboratories, Keele University, ST5 5BG, UK}
\affil[3]{\scriptsize Jodrell Bank Centre for Astrophysics, Alan Turing Building, University of Manchester, M13 9PL, UK}

\affil[4]{\scriptsize Iranian National Observatory, Institute for Research in Fundamental Sciences (IPM), Tehran, Iran}

\begin{document}
\pagestyle{empty}
\newpage
\pagestyle{fancy}
\label{firstpage}
\date{}
\maketitle

\begin{abstract}
Differences in metallicity between the Large Magellanic Cloud (LMC) and the Small Magellanic Cloud (SMC) offer an opportunity to examine whether environmental metallicity affects the performance of machine learning models in classifying dusty stellar sources. The five stellar classes studied include young stellar objects (YSOs), red supergiants (RSGs), post-asymptotic giant branch stars (PAGBs), and oxygen- and carbon-rich asymptotic giant branch stars (OAGBs and CAGBs), which are key phases of stellar evolution involved in dust production. Using spectroscopically labeled data from the Surveying the Agents of Galaxy Evolution (SAGE) project, we trained and evaluated a probabilistic random forest (PRF) classifier with four approaches: (1) separate training on LMC and SMC, including all five classes, (2) excluding the underpopulated PAGB class, (3) combined LMC and SMC datasets, and (4) cross-galaxy training and testing. The model achieved 93\% accuracy on the SMC and 88\% on the LMC across all five classes. In the SMC, PAGB sources were misclassified as YSOs, mainly because of their small sample size (4 objects). When PAGB was excluded, both the LMC and the SMC reached 92\% accuracy. A combined dataset produced the same accuracy, and cross-galaxy training yielded similar results, indicating that metallicity does not significantly impact model performance. A comparison of absolute CMDs for the LMC and SMC confirms their similarity in stellar populations. These findings suggest that environmental metallicity has little effect on ML-based classification of dusty stellar sources, supporting the use of combined datasets and cross-galaxy models in low-metallicity environments.\\

\end{abstract}

%%%%%%% Introduction %%%%%%%%
\section {Introduction}
The Large Magellanic Cloud (LMC) and Small Magellanic Cloud (SMC) are nearby dwarf galaxies at distances of approximately 50 and 60 kpc, with metallicities of about 0.5 $Z_{\odot}$ and 0.2 $Z_{\odot}$, respectively \citep{1992ApJ-metalicity-russel, Pietrzy-lmc-2013Natur.495...76P, LMC-SMC-2009A&A...496..399S}. Their proximity and contrasting chemical compositions make them excellent laboratories for investigating the relationship between stellar evolution, dust formation, and environmental metallicity \citep{Ruffle2015}.

Dusty stellar sources, including young stellar objects (YSOs) and evolved stars—oxygen- and carbon-rich asymptotic giant branch stars (OAGBs, CAGBs), red supergiants (RSGs), and post-asymptotic giant branch stars (PAGBs)—represent key phases in the stellar life cycle. These objects return heavy elements to the interstellar medium, shaping the dust content and chemical enrichment of galaxies \citep{Boyer2011b, 2018A&AR-massloss-agb-hofner}. Because of their brightness and strong infrared excesses, these populations are particularly well traced in the Magellanic Clouds.

Traditional classification of dusty stellar populations has relied primarily on photometric methods, such as color–magnitude diagrams (CMDs) and spectral energy distributions (SEDs). While such techniques are efficient for large samples, they can be ambiguous due to overlapping photometric signatures among different stellar classes. Spectroscopic observations, on the other hand, provide a more reliable basis for classification but remain limited in number. The Surveying the Agents of Galaxy Evolution (SAGE) and its spectroscopic follow-up SAGE-Spec surveys \citep{Kemper2010, Wood2011, Ruffle2015, Jones2017,2006AJ-Meixner} have provided infrared photometry and spectroscopically confirmed dusty stellar sources in the Magellanic Clouds, enabling direct machine-learning studies of evolved stars and YSOs in both galaxies.

Machine learning (ML) methods have become essential tools for identifying and classifying stellar sources from multiwavelength data \citep{2019arXiv190407248B}. Among these, the Probabilistic Random Forest (PRF) classifier \citep{2019AJ....157...16R, 2022MNRAS.517..140K,Ghaziasgar2024-BAO,2025-Pennock} has proven particularly effective for handling uncertainties and imbalanced datasets. Previous studies have shown that supervised ML models trained on spectroscopic labels can achieve accuracies of around 90\% in categorizing dusty stellar sources \citep{Ghaziasgar2024-BAO}.

However, differences in metallicity between the LMC and SMC may influence the photometric and spectroscopic properties of dusty stellar sources. In this study, we investigate whether the metallicity contrast between the two galaxies affects the performance of machine-learning classification of these objects, using spectroscopically labeled datasets from the Magellanic Clouds.

%%%%%
\section{Data and Model}
%%%%
The dataset is obtained from the SAGE and SAGE-Spec surveys \citep{Kemper2010, Wood2011, Ruffle2015, Jones2017,2006AJ-Meixner}, which provide multiwavelength photometry and spectroscopically classified dusty stellar sources in the Magellanic Clouds. The combined spectroscopic catalog includes about 618 sources classified into five stellar categories: young stellar objects (YSOs), oxygen- and carbon-rich asymptotic giant branch stars (OAGBs and CAGBs), red supergiants (RSGs), and post-asymptotic giant branch stars (PAGBs).

The same 12 near- and mid-infrared filters used in these studies \citep{Ghaziasgar2024-BAO, Ghaziasgar2025-ApJ, Ghaziasgar-IAUS2025,Ghaziasgar2025-IAUS-Greece} were adopted here, as they effectively trace both photospheric and circumstellar dust emission. These features are particularly sensitive to dust and molecular absorption bands, making them suitable for identifying dusty stellar classes.

The full sample was naturally divided into two subsets corresponding to sources located in the Large and Small Magellanic Clouds, allowing a direct comparison between environments with different metallicities. The LMC subset contains 486 sources, while the SMC contains 132, reflecting the intrinsic population imbalance between the two galaxies. Due to the small number of PAGB stars in the SMC (only four objects), the Synthetic Minority Oversampling Technique (SMOTE; \citealt{2011arXiv1106.1813C}) could not be applied, since it requires at least six samples per class to generate synthetic data points.

The PRF classifier, previously identified as the most accurate model in \citet{Ghaziasgar2024-BAO}, was adopted to evaluate classification performance. To assess the potential impact of metallicity in a systematic manner, we designed four experimental configurations: (1) training the model separately on the LMC and SMC datasets, (2) repeating the analysis after removing the sparsely populated PAGB class, (3) training and testing on the combined LMC and SMC catalog, and (4) cross-galaxy training and testing. Model performance was evaluated using precision, recall, F1-score, and accuracy, with confusion matrices used to visualize class-level outcomes.

\section{Results}

\begin{figure*}[h!]
    \centering
    \includegraphics[width=0.42\linewidth]{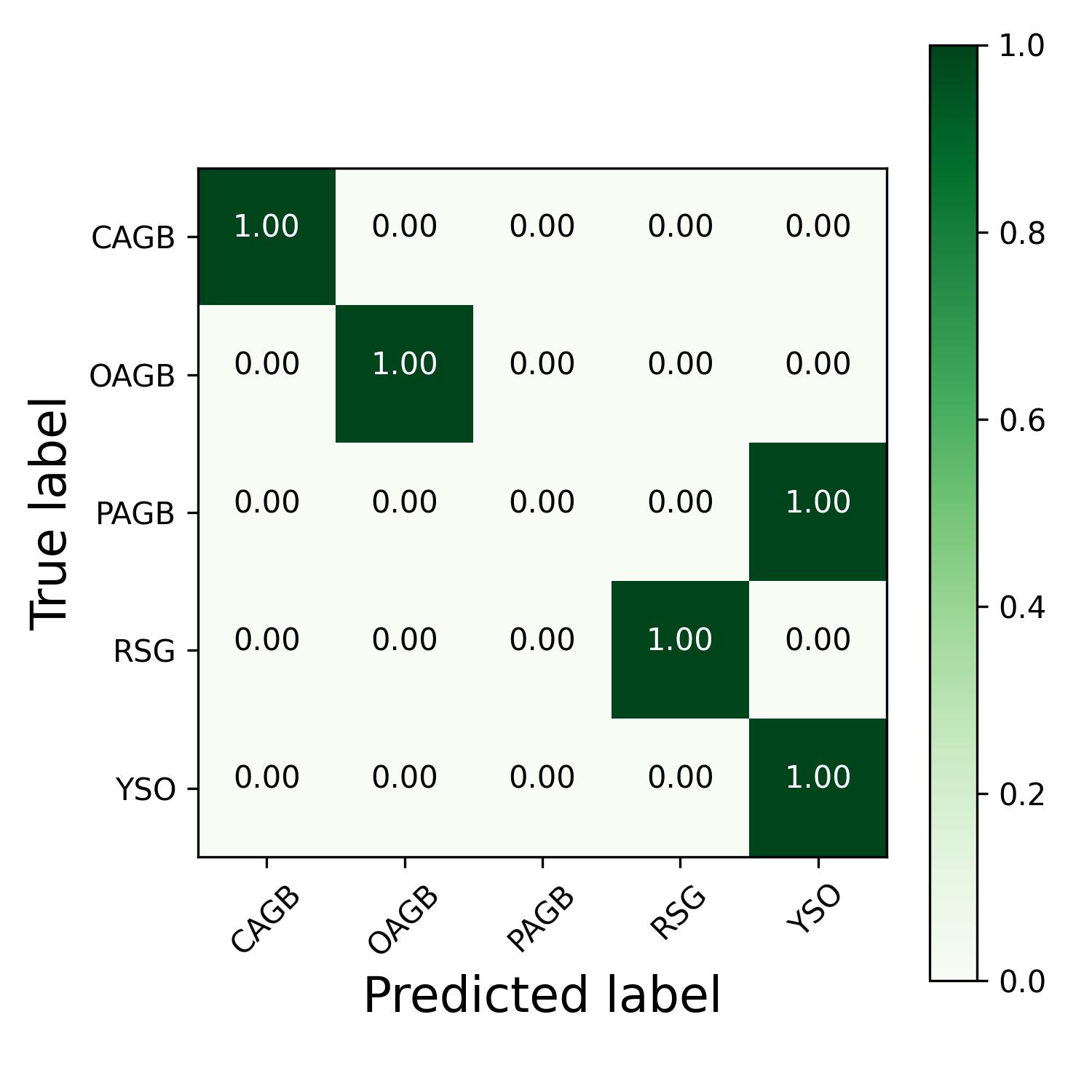}
    \hspace{0.06\linewidth}
    \includegraphics[width=0.42\linewidth]{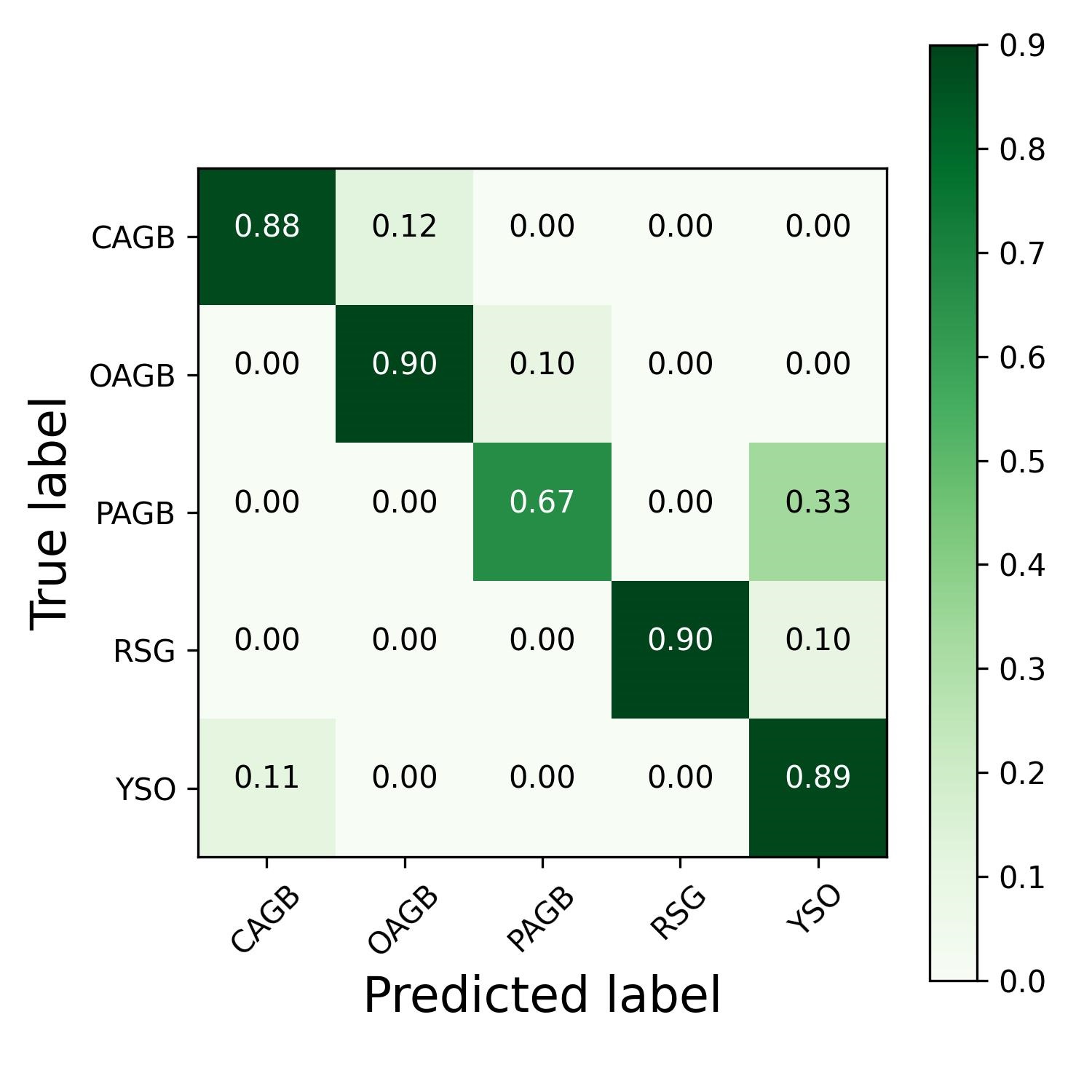}
    \caption{Confusion matrices for PRF classifier applied separately to the SMC (left) and LMC (right) using five stellar classes.}
    \label{fig:CM-5class}
\end{figure*}

To test whether metallicity influences classification performance, we first trained the PRF model separately on the LMC and SMC catalogs while keeping all five stellar classes. The SMC experiment achieved an overall accuracy of 93\%, correctly identifying four classes; however, PAGB sources were not recovered and were assigned mainly to the YSO category due to their extremely small sample size (four objects). The limited size of the SMC set (132 objects) also prevented the use of SMOTE. These results are presented in the classification reports (Tables~\ref{tab:classification_report-SMC}, \ref{tab:classification_report-LMC}) and the corresponding confusion matrices (Fig.~\ref{fig:CM-5class}).

\begin{table}[h!]
    \centering
    \caption{Classification report for the SMC catalog (five classes).}
    \label{tab:classification_report-SMC}
    \begin{tabular}{lccc}
        \hline
        Class & Precision & Recall & F1-Score \\
        \hline
        CAGB & 1.00 & 1.00 & 1.00 \\
        OAGB & 1.00 & 1.00 & 1.00 \\
        PAGB & 0.00 & 0.00 & 0.00 \\
        RSG  & 1.00 & 1.00 & 1.00 \\
        YSO  & 0.80 & 1.00 & 0.89 \\
        \hline
        accuracy & & & 0.93 \\
        macro avg & 0.76 & 0.80 & 0.78 \\
        weighted avg & 0.88 & 0.93 & 0.90 \\
        \hline
    \end{tabular}
\end{table}

\begin{table}[h!]
    \centering
    \caption{Classification report for the LMC catalog (five classes).}
    \label{tab:classification_report-LMC}
    \begin{tabular}{lccc}
        \hline
        Class & Precision & Recall & F1-Score \\
        \hline
        CAGB & 0.94 & 0.88 & 0.91 \\
        OAGB & 0.72 & 0.90 & 0.86 \\
        PAGB & 0.67 & 0.67 & 0.67 \\
        RSG  & 1.00 & 0.90 & 0.95 \\
        YSO  & 0.80 & 0.80 & 0.88 \\
        \hline
        accuracy & & & 0.88 \\
        macro avg & 0.84 & 0.85 & 0.84 \\
        weighted avg & 0.80 & 0.88 & 0.88 \\
        \hline
    \end{tabular}
\end{table}

\FloatBarrier
\clearpage

\begin{figure*}[h!]
\centering
\includegraphics[width=\textwidth]{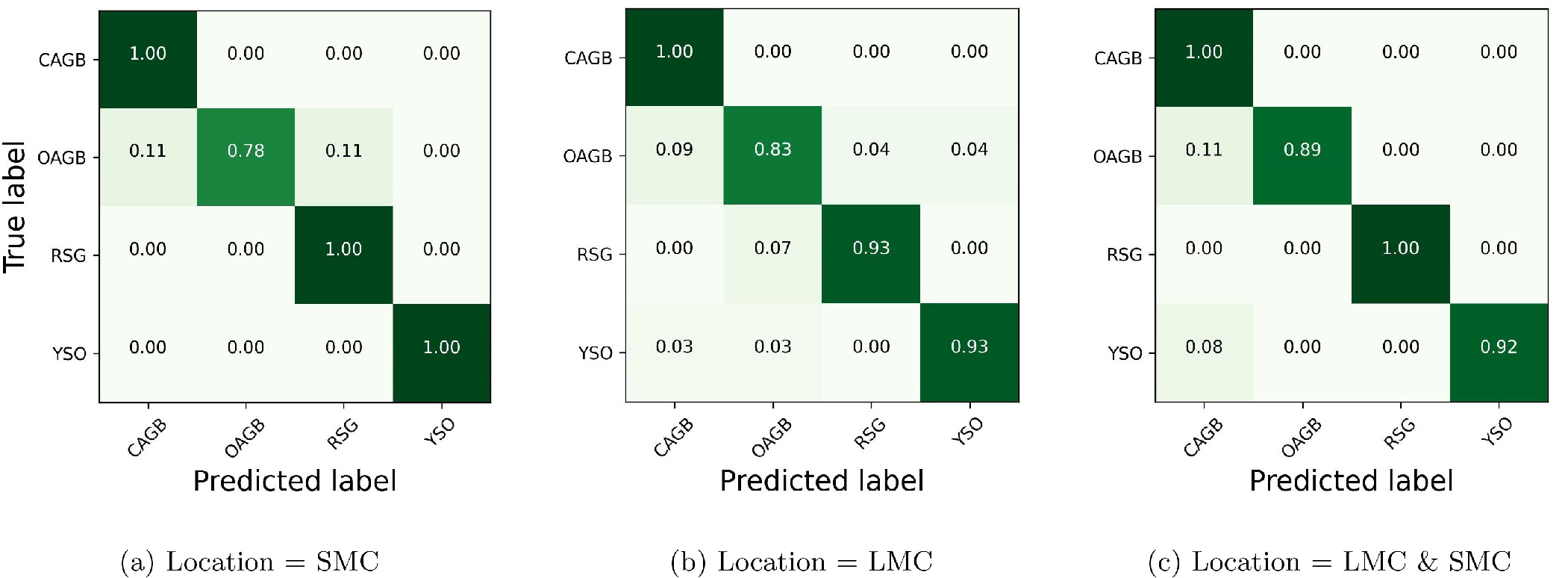}
\caption{Confusion matrices for four-class PRF experiments: SMC-only (left), LMC-only (middle), and combined LMC\&SMC (right).}
\label{fig:CM-Metallicity-4Class}
\end{figure*}

To remove the effect of the sparsely populated PAGB class, we repeated the analysis using only the four well-represented classes. In this configuration, both the LMC and SMC achieved an accuracy of 92\%, confirming that PAGB imbalance was the primary limiting factor. The reports and confusion matrices are shown in Tables~\ref{tab:classification_report_SMC_4class}, \ref{tab:classification_report_LMC_4class} and Fig.~\ref{fig:CM-Metallicity-4Class}. The similar diagonal patterns demonstrate that the main stellar groups are consistently learned, indicating that metallicity alone does not introduce measurable differences.

\begin{table}[h!]
    \centering
    \caption{Classification report for the SMC catalog (four classes).}
    \label{tab:classification_report_SMC_4class}
    \begin{tabular}{lccc}
        \hline
        Class & Precision & Recall & F1-Score \\
        \hline
        CAGB & 0.75 & 1.00 & 0.86 \\
        OAGB & 0.86 & 0.78 & 0.88 \\
        RSG  & 0.67 & 1.00 & 0.80 \\
        YSO  & 1.00 & 1.00 & 1.00 \\
        \hline
        accuracy & & & 0.92 \\
        macro avg & 0.85 & 0.94 & 0.88 \\
        weighted avg & 0.95 & 0.92 & 0.92 \\
        \hline
    \end{tabular}
\end{table}

\begin{table}[h!]
    \centering
    \caption{Classification report for the LMC catalog (four classes).}
    \label{tab:classification_report_LMC_4class}
    \begin{tabular}{lccc}
        \hline
        Class & Precision & Recall & F1-Score \\
        \hline
        CAGB & 0.88 & 1.00 & 0.94 \\
        OAGB & 0.90 & 0.83 & 0.86 \\
        RSG  & 0.93 & 0.93 & 0.93 \\
        YSO  & 0.97 & 0.93 & 0.95 \\
        \hline
        accuracy & & & 0.92 \\
        macro avg & 0.92 & 0.92 & 0.92 \\
        weighted avg & 0.92 & 0.92 & 0.92 \\
        \hline
    \end{tabular}
\end{table}

\FloatBarrier
\clearpage

\begin{figure}[h!]
  \centering
  \includegraphics[width=0.5\linewidth]{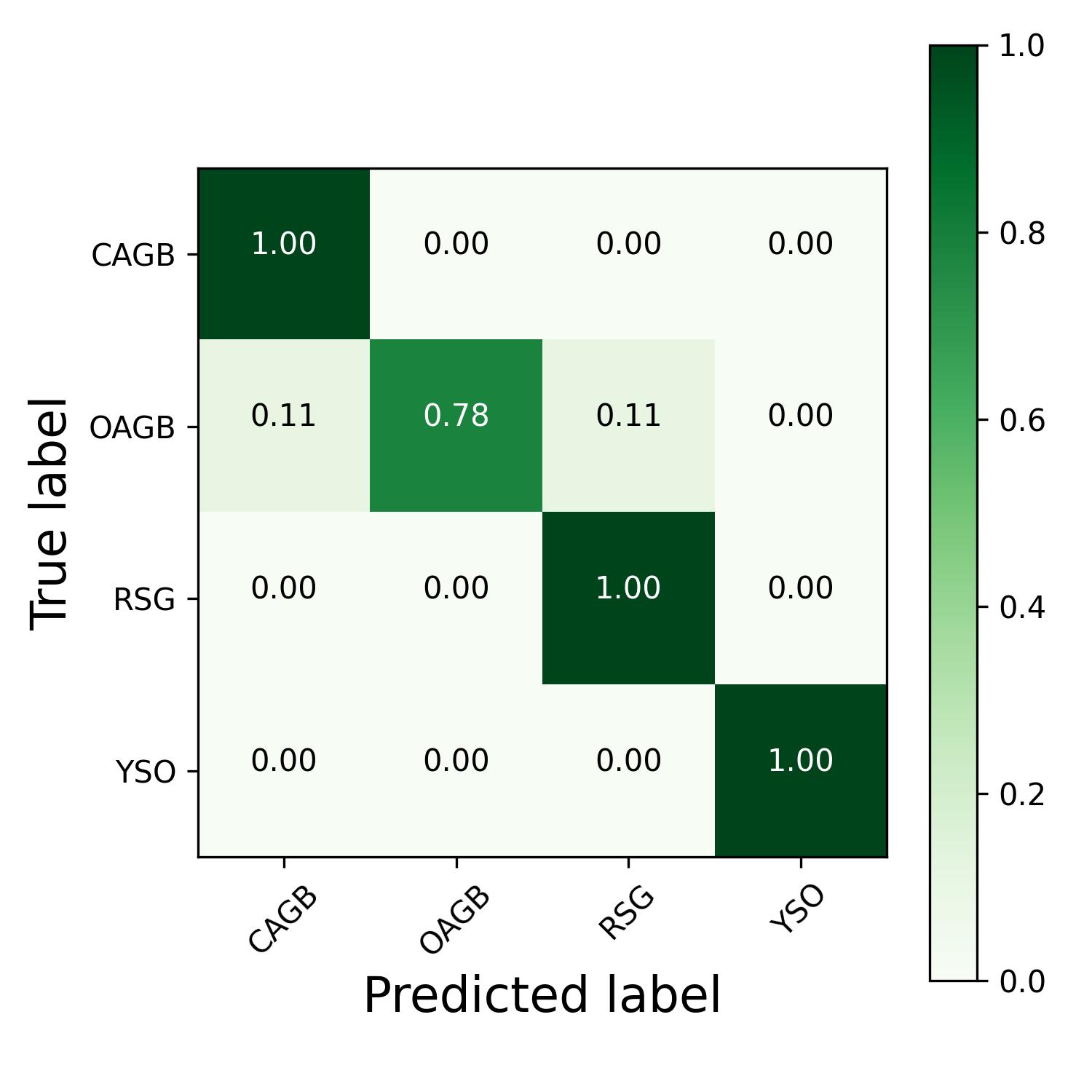}
  \caption{Confusion matrix for the model trained on LMC and tested on SMC (four classes).}
  \label{fig:CM-cross}
\end{figure}

Finally, we trained a model on the LMC sources and applied it to the SMC data. The accuracy again reached 92\%, demonstrating that the model generalizes well across different metallicity environments. The classification report and confusion matrix are shown in Table~\ref{tab:classification_report_LMC_Train_SMC_test_4class} and Fig.~\ref{fig:CM-cross}.

\begin{table}[h!]
    \centering
    \caption{Classification report for the model trained on LMC and tested on SMC (four classes).}
    \label{tab:classification_report_LMC_Train_SMC_test_4class}
    \begin{tabular}{lccc}
        \hline
        Class & Precision & Recall & F1-Score \\
        \hline
        CAGB & 0.75 & 1.00 & 0.86 \\
        OAGB & 0.86 & 0.78 & 0.88 \\
        RSG  & 0.67 & 1.00 & 0.80 \\
        YSO  & 1.00 & 1.00 & 1.00 \\
        \hline
        accuracy & & & 0.92 \\
        macro avg & 0.85 & 0.94 & 0.88 \\
        weighted avg & 0.95 & 0.92 & 0.92 \\
        \hline
    \end{tabular}
\end{table}

In summary, these results show that the metallicity contrast between the LMC and SMC has no measurable effect on the PRF classifier's performance. Across all four experimental configurations, including separate training, the runs without the PAGB class, the combined datasets, and the cross-galaxy tests, the model reaches nearly identical accuracies and produces similarly structured confusion matrices. This demonstrates that a classifier trained in one low-metallicity environment remains reliable when applied to the other. In practice, this allows the use of unified training sets, reduces dependence on limited spectroscopic samples, and supports extending machine-learning classification to other low-metallicity dwarf galaxies in the Local Group.

\FloatBarrier
%%%%%%%%%%%%%%%%%%%%%%%%%%%%%%%%%%
\clearpage % To force this stuff to happen by this point in the text, otherwise these will probably end up after the references.

\section*{\small Acknowledgements}
\scriptsize{The authors thank the School of Astronomy at the Institute for Research in Fundamental Sciences (IPM) and the Iranian National Observatory (INO) for supporting this
research. S. Ghaziasgar is grateful for the support of the Byurakan Astrophysical Observatory (BAO).}

\scriptsize
\bibliographystyle{ComBAO}
\nocite{*}
\bibliography{references}

@ARTICLE{Ruffle2015,
       author = {{Ruffle}, Paul M.~E. and {Kemper}, F. and {Jones}, O.~C. and {Sloan}, G.~C. and {Kraemer}, K.~E. and {Woods}, Paul M. and {Boyer}, M.~L. and {Srinivasan}, S. and {Antoniou}, V. and {Lagadec}, E. and {Matsuura}, M. and {McDonald}, I. and {Oliveira}, J.~M. and {Sargent}, B.~A. and {Sewi{\l}o}, M. and {Szczerba}, R. and {van Loon}, J. Th. and {Volk}, K. and {Zijlstra}, A.~A.},
        title = "{Spitzer infrared spectrograph point source classification in the Small Magellanic Cloud}",
      journal = {\mnras},
         year = 2015,
        month = aug,
       volume = {451},
       number = {4},
        pages = {3504-3536},
          doi = {10.1093/mnras/stv1106},
archivePrefix = {arXiv},
       eprint = {1505.04499},
 primaryClass = {astro-ph.SR},
       adsurl = {https://ui.adsabs.harvard.edu/abs/2015MNRAS.451.3504R}
}

@ARTICLE{2022MNRAS.517..140K,
       author = {{Kinson}, David A. and {Oliveira}, Joana M. and {van Loon}, Jacco Th},
        title = "{Massive young stellar objects in the Local Group spiral galaxy M 33 identified using machine learning}",
      journal = {\mnras},
         year = 2022,
        month = nov,
       volume = {517},
       number = {1},
        pages = {140-160},
          doi = {10.1093/mnras/stac2692},
archivePrefix = {arXiv},
       eprint = {2209.10371},
 primaryClass = {astro-ph.GA},
       adsurl = {https://ui.adsabs.harvard.edu/abs/2022MNRAS.517..140K}
}

@ARTICLE{2019arXiv190407248B,
       author = {{Baron}, Dalya},
        title = "{Machine Learning in Astronomy: a practical overview}",
      journal = {arXiv e-prints},
         year = 2019,
        month = apr,
          eid = {arXiv:1904.07248},
        pages = {arXiv:1904.07248},
          doi = {10.48550/arXiv.1904.07248},
archivePrefix = {arXiv},
       eprint = {1904.07248},
 primaryClass = {astro-ph.IM},
       adsurl = {https://ui.adsabs.harvard.edu/abs/2019arXiv190407248B}
}

@ARTICLE{2019AJ....157...16R,
       author = {{Reis}, Itamar and {Baron}, Dalya and {Shahaf}, Sahar},
        title = "{Probabilistic Random Forest: A Machine Learning Algorithm for Noisy Data Sets}",
      journal = {\aj},
         year = 2019,
        month = jan,
       volume = {157},
       number = {1},
          eid = {16},
        pages = {16},
          doi = {10.3847/1538-3881/aaf101},
archivePrefix = {arXiv},
       eprint = {1811.05994},
 primaryClass = {astro-ph.IM},
       adsurl = {https://ui.adsabs.harvard.edu/abs/2019AJ....157...16R}
}

@ARTICLE{2011arXiv1106.1813C,
       author = {{Chawla}, N.~V. and {Bowyer}, K.~W. and {Hall}, L.~O. and {Kegelmeyer}, W.~P.},
        title = "{SMOTE: Synthetic Minority Over-sampling Technique}",
      journal = {arXiv e-prints},
         year = 2011,
        month = jun,
          eid = {arXiv:1106.1813},
        pages = {arXiv:1106.1813},
          doi = {10.48550/arXiv.1106.1813},
archivePrefix = {arXiv},
       eprint = {1106.1813},
 primaryClass = {cs.AI},
       adsurl = {https://ui.adsabs.harvard.edu/abs/2011arXiv1106.1813C}
}

@ARTICLE{2006AJ-Meixner,
       author = {{Meixner}, Margaret and {Gordon}, Karl D. and {Indebetouw}, Remy and {Hora}, Joseph L. and {Whitney}, Barbara and {Blum}, Robert and {Reach}, William and {Bernard}, Jean-Philippe and {Meade}, Marilyn and {Babler}, Brian and {Engelbracht}, Charles W. and {For}, Bi-Qing and {Misselt}, Karl and {Vijh}, Uma and {Leitherer}, Claus and {Cohen}, Martin and {Churchwell}, Ed B. and {Boulanger}, Francois and {Frogel}, Jay A. and {Fukui}, Yasuo and {Gallagher}, Jay and {Gorjian}, Varoujan and {Harris}, Jason and {Kelly}, Douglas and {Kawamura}, Akiko and {Kim}, SoYoung and {Latter}, William B. and {Madden}, Suzanne and {Markwick-Kemper}, Ciska and {Mizuno}, Akira and {Mizuno}, Norikazu and {Mould}, Jeremy and {Nota}, Antonella and {Oey}, M.~S. and {Olsen}, Knut and {Onishi}, Toshikazu and {Paladini}, Roberta and {Panagia}, Nino and {Perez-Gonzalez}, Pablo and {Shibai}, Hiroshi and {Sato}, Shuji and {Smith}, Linda and {Staveley-Smith}, Lister and {Tielens}, A.~G.~G.~M. and {Ueta}, Toshiya and {van Dyk}, Schuyler and {Volk}, Kevin and {Werner}, Michael and {Zaritsky}, Dennis},
        title = "{Spitzer Survey of the Large Magellanic Cloud: Surveying the Agents of a Galaxy's Evolution (SAGE). I. Overview and Initial Results}",
      journal = {\aj},
         year = 2006,
        month = dec,
       volume = {132},
       number = {6},
        pages = {2268-2288},
          doi = {10.1086/508185},
archivePrefix = {arXiv},
       eprint = {astro-ph/0606356},
 primaryClass = {astro-ph},
       adsurl = {https://ui.adsabs.harvard.edu/abs/2006AJ....132.2268M}
}

@ARTICLE{Boyer2011b,
       author = {{Boyer}, Martha L. and {Srinivasan}, Sundar and {van Loon}, Jacco Th. and {McDonald}, Iain and {Meixner}, Margaret and {Zaritsky}, Dennis and {Gordon}, Karl D. and {Kemper}, F. and {Babler}, Brian and {Block}, Miwa and {Bracker}, Steve and {Engelbracht}, Charles W. and {Hora}, Joe and {Indebetouw}, Remy and {Meade}, Marilyn and {Misselt}, Karl and {Robitaille}, Thomas and {Sewi{\l}o}, Marta and {Shiao}, Bernie and {Whitney}, Barbara},
        title = "{Surveying the Agents of Galaxy Evolution in the Tidally Stripped, Low Metallicity Small Magellanic Cloud (SAGE-SMC). II. Cool Evolved Stars}",
      journal = {\aj},
         year = 2011,
        month = oct,
       volume = {142},
       number = {4},
          eid = {103},
        pages = {103},
          doi = {10.1088/0004-6256/142/4/103},
archivePrefix = {arXiv},
       eprint = {1106.5026},
 primaryClass = {astro-ph.SR},
       adsurl = {https://ui.adsabs.harvard.edu/abs/2011AJ....142..103B}
}

@ARTICLE{1992ApJ-metalicity-russel,
       author = {{Russell}, Stephen C. and {Dopita}, Michael A.},
        title = "{Abundances of the Heavy Elements in the Magellanic Clouds. III. Interpretation of Results}",
      journal = {\apj},
         year = 1992,
        month = jan,
       volume = {384},
        pages = {508},
          doi = {10.1086/170893},
       adsurl = {https://ui.adsabs.harvard.edu/abs/1992ApJ...384..508R}
}

@ARTICLE{Kemper2010,
       author = {{Kemper}, F. and {Woods}, Paul M. and {Antoniou}, V. and {Bernard}, J.-P. and {Blum}, R.~D. and {Boyer}, M.~L. and {Chan}, J. and {Chen}, C.-H.~R. and {Cohen}, M. and {Dijkstra}, C. and {Engelbracht}, C. and {Galametz}, M. and {Galliano}, F. and {Gielen}, C. and {Gordon}, Karl D. and {Gorjian}, V. and {Harris}, J. and {Hony}, S. and {Hora}, J.~L. and {Indebetouw}, R. and {Jones}, O. and {Kawamura}, A. and {Lagadec}, E. and {Lawton}, B. and {Leisenring}, J.~M. and {Madden}, S.~C. and {Marengo}, M. and {Matsuura}, M. and {McDonald}, I. and {McGuire}, C. and {Meixner}, M. and {Mulia}, A.~J. and {O'Halloran}, B. and {Oliveira}, J.~M. and {Paladini}, R. and {Paradis}, D. and {Reach}, W.~T. and {Rubin}, D. and {Sandstrom}, K. and {Sargent}, B.~A. and {Sewilo}, M. and {Shiao}, B. and {Sloan}, G.~C. and {Speck}, A.~K. and {Srinivasan}, S. and {Szczerba}, R. and {Tielens}, A.~G.~G.~M. and {van Aarle}, E. and {Van Dyk}, S.~D. and {van Loon}, J. Th. and {Van Winckel}, H. and {Vijh}, Uma P. and {Volk}, K. and {Whitney}, B.~A. and {Wilkins}, A.~N. and {Zijlstra}, A.~A.},
        title = "{The SAGE-Spec Spitzer Legacy Program: The Life Cycle of Dust and Gas in the Large Magellanic Cloud}",
      journal = {\pasp},
         year = 2010,
        month = jun,
       volume = {122},
       number = {892},
        pages = {683},
          doi = {10.1086/653438},
archivePrefix = {arXiv},
       eprint = {1004.1142},
 primaryClass = {astro-ph.GA},
       adsurl = {https://ui.adsabs.harvard.edu/abs/2010PASP..122..683K}
}

@ARTICLE{Jones2017,
       author = {{Jones}, O.~C. and {Woods}, P.~M. and {Kemper}, F. and {Kraemer}, K.~E. and {Sloan}, G.~C. and {Srinivasan}, S. and {Oliveira}, J.~M. and {van Loon}, J. Th. and {Boyer}, M.~L. and {Sargent}, B.~A. and {McDonald}, I. and {Meixner}, M. and {Zijlstra}, A.~A. and {Ruffle}, P.~M.~E. and {Lagadec}, E. and {Pauly}, T. and {Sewi{\l}o}, M. and {Clayton}, G.~C. and {Volk}, K.},
        title = "{The SAGE-Spec Spitzer Legacy program: the life-cycle of dust and gas in the Large Magellanic Cloud. Point source classification - III}",
      journal = {\mnras},
         year = 2017,
        month = sep,
       volume = {470},
       number = {3},
        pages = {3250-3282},
          doi = {10.1093/mnras/stx1101},
archivePrefix = {arXiv},
       eprint = {1705.02709},
 primaryClass = {astro-ph.SR},
       adsurl = {https://ui.adsabs.harvard.edu/abs/2017MNRAS.470.3250J}
}

@ARTICLE{Wood2011,
       author = {{Woods}, Paul M. and {Oliveira}, J.~M. and {Kemper}, F. and {van Loon}, J. Th. and {Sargent}, B.~A. and {Matsuura}, M. and {Szczerba}, R. and {Volk}, K. and {Zijlstra}, A.~A. and {Sloan}, G.~C. and {Lagadec}, E. and {McDonald}, I. and {Jones}, O. and {Gorjian}, V. and {Kraemer}, K.~E. and {Gielen}, C. and {Meixner}, M. and {Blum}, R.~D. and {Sewi{\l}o}, M. and {Riebel}, D. and {Shiao}, B. and {Chen}, C.-H.~R. and {Boyer}, M.~L. and {Indebetouw}, R. and {Antoniou}, V. and {Bernard}, J.-P. and {Cohen}, M. and {Dijkstra}, C. and {Galametz}, M. and {Galliano}, F. and {Gordon}, Karl D. and {Harris}, J. and {Hony}, S. and {Hora}, J.~L. and {Kawamura}, A. and {Lawton}, B. and {Leisenring}, J.~M. and {Madden}, S. and {Marengo}, M. and {McGuire}, C. and {Mulia}, A.~J. and {O'Halloran}, B. and {Olsen}, K. and {Paladini}, R. and {Paradis}, D. and {Reach}, W.~T. and {Rubin}, D. and {Sandstrom}, K. and {Soszy{\'n}ski}, I. and {Speck}, A.~K. and {Srinivasan}, S. and {Tielens}, A.~G.~G.~M. and {van Aarle}, E. and {van Dyk}, S.~D. and {van Winckel}, H. and {Vijh}, Uma P. and {Whitney}, B. and {Wilkins}, A.~N.},
        title = "{The SAGE-Spec Spitzer Legacy programme: the life-cycle of dust and gas in the Large Magellanic Cloud - Point source classification I}",
      journal = {\mnras},
         year = 2011,
        month = mar,
       volume = {411},
       number = {3},
        pages = {1597-1627},
          doi = {10.1111/j.1365-2966.2010.17794.x},
archivePrefix = {arXiv},
       eprint = {1009.5929},
 primaryClass = {astro-ph.GA},
       adsurl = {https://ui.adsabs.harvard.edu/abs/2011MNRAS.411.1597W}
}

@ARTICLE{Ghaziasgar2025-ApJ,
       author = {{Ghaziasgar}, Sepideh and {Abdollahi}, Mahdi and {Javadi}, Atefeh and {van Loon}, Jacco Th. and {McDonald}, Iain and {Oliveira}, Joana and {Masoudnezhad}, Amirhossein and {Khosroshahi}, Habib G. and {Foing}, Bernard H. and {Fazel Hesar}, Fatemeh},
        title = "{Dusty Stellar Source Classification by Implementing Machine Learning Methods Based on Spectroscopic Observations in the Magellanic Clouds}",
      journal = {\apj},
         year = 2025,
        month = jun,
       volume = {986},
       number = {2},
          eid = {168},
        pages = {168},
          doi = {10.3847/1538-4357/adceeb},
archivePrefix = {arXiv},
       eprint = {2504.14332},
 primaryClass = {astro-ph.GA},
       adsurl = {https://ui.adsabs.harvard.edu/abs/2025ApJ...986..168G}
}

@INPROCEEDINGS{Ghaziasgar-IAUS2025,
       author = {{Ghaziasgar}, Sepideh and {Masoudnezhad}, Amirhossein and {Javadi}, Atefeh and {van Loon}, Jacco Th. and {Khosroshahi}, Habib G. and {Khosravaninezhad}, Negin},
        title = "{Spectral identification and classification of dusty stellar sources using spectroscopic and multiwavelength observations through machine learning}",
    booktitle = {IAU Symposium},
         year = 2025,
       editor = {{McIver}, J. and {Mahabal}, A. and {Fluke}, C.},
       series = {IAU Symposium},
       volume = {19},
        month = aug,
        pages = {67-72},
          doi = {10.1017/S1743921323001308},
archivePrefix = {arXiv},
       eprint = {2211.03403},
 primaryClass = {astro-ph.GA},
       adsurl = {https://ui.adsabs.harvard.edu/abs/2025IAUS..368...67G}
}

@ARTICLE{Ghaziasgar2024-BAO,
       author = {{Ghaziasgar}, S. and {Abdollahi}, M. and {Javadi}, A. and {van Loo}, J. Th. and {McDonald}, I. and {Oliveira}, J. and {Khosroshahi}, H.~G.},
        title = "{Machine Learning Classification of Young Stellar Objects and Evolved Stars in the Magellanic Clouds Using the Probabilistic Random Forest Classifier}",
      journal = {Communications of the Byurakan Astrophysical Observatory},
         year = 2024,
        month = dec,
       volume = {71},
        pages = {377-382},
          doi = {10.52526/25792776-24.71.2-377},
archivePrefix = {arXiv},
       eprint = {2504.14242},
 primaryClass = {astro-ph.GA},
       adsurl = {https://ui.adsabs.harvard.edu/abs/2024CoBAO..71..377G}
}

@ARTICLE{Ghaziasgar2025-IAUS-Greece,
       author = {{Ghaziasgar}, Sepideh and {Abdollahi}, Mahdi and {Javadi}, Atefeh and {van Loon}, Jacco Th. and {McDonald}, Iain and {Oliveira}, Joana and {Khosroshahi}, Habib G.},
        title = "{Comparison of Photometric and Spectroscopic Labels in Classifying Dusty Stellar Sources Using Machine Learning in the Magellanic Clouds}",
      journal = {arXiv e-prints},
         year = 2025,
        month = sep,
          eid = {arXiv:2509.05531},
        pages = {arXiv:2509.05531},
          doi = {10.48550/arXiv.2509.05531},
archivePrefix = {arXiv},
       eprint = {2509.05531},
 primaryClass = {astro-ph.GA},
       adsurl = {https://ui.adsabs.harvard.edu/abs/2025arXiv250905531G}
}

@ARTICLE{2018A&AR-massloss-agb-hofner,
       author = {{H{\"o}fner}, Susanne and {Olofsson}, Hans},
        title = "{Mass loss of stars on the asymptotic giant branch. Mechanisms, models and measurements}",
      journal = {\aapr},
         year = 2018,
        month = jan,
       volume = {26},
       number = {1},
          eid = {1},
        pages = {1},
          doi = {10.1007/s00159-017-0106-5},
       adsurl = {https://ui.adsabs.harvard.edu/abs/2018A&ARv..26....1H}
}

@ARTICLE{LMC-SMC-2009A&A...496..399S,
       author = {{Subramanian}, S. and {Subramaniam}, A.},
        title = "{Depth estimation of the Large and Small Magellanic Clouds}",
      journal = {\aap},
         year = 2009,
        month = mar,
       volume = {496},
       number = {2},
        pages = {399-412},
          doi = {10.1051/0004-6361/200811029},
archivePrefix = {arXiv},
       eprint = {0809.4362},
 primaryClass = {astro-ph},
       adsurl = {https://ui.adsabs.harvard.edu/abs/2009A&A...496..399S}
}

@ARTICLE{Pietrzy-lmc-2013Natur.495...76P,
       author = {{Pietrzy{\'n}ski}, G. and {Graczyk}, D. and {Gieren}, W. and {Thompson}, I.~B. and {Pilecki}, B. and {Udalski}, A. and {Soszy{\'n}ski}, I. and {Koz{\l}owski}, S. and {Konorski}, P. and {Suchomska}, K. and {Bono}, G. and {Moroni}, P.~G. Prada and {Villanova}, S. and {Nardetto}, N. and {Bresolin}, F. and {Kudritzki}, R.~P. and {Storm}, J. and {Gallenne}, A. and {Smolec}, R. and {Minniti}, D. and {Kubiak}, M. and {Szyma{\'n}ski}, M.~K. and {Poleski}, R. and {Wyrzykowski}, {\L}. and {Ulaczyk}, K. and {Pietrukowicz}, P. and {G{\'o}rski}, M. and {Karczmarek}, P.},
        title = "{An eclipsing-binary distance to the Large Magellanic Cloud accurate to two per cent}",
      journal = {\nat},
         year = 2013,
        month = mar,
       volume = {495},
       number = {7439},
        pages = {76-79},
          doi = {10.1038/nature11878},
archivePrefix = {arXiv},
       eprint = {1303.2063},
 primaryClass = {astro-ph.GA},
       adsurl = {https://ui.adsabs.harvard.edu/abs/2013Natur.495...76P}
}

@ARTICLE{2025-Pennock,
       author = {{Pennock}, Clara M. and {van Loon}, Jacco Th and {Cioni}, Maria-Rosa L. and {Maitra}, Chandreyee and {Oliveira}, Joana M. and {Craig}, Jessica E.~M. and {Ivanov}, Valentin D. and {Aird}, James and {Anih}, Joy O. and {Cross}, Nicholas J.~G. and {Dresbach}, Francesca and {de Grijs}, Richard and {Groenewegen}, Martin A.~T.},
        title = "{The VMC Survey - LI. Classifying extragalactic sources using a probabilistic random forest supervised machine learning algorithm}",
      journal = {\mnras},
         year = 2025,
        month = feb,
       volume = {537},
       number = {2},
        pages = {1028-1055},
          doi = {10.1093/mnras/staf080},
archivePrefix = {arXiv},
       eprint = {2501.08196},
 primaryClass = {astro-ph.GA},
       adsurl = {https://ui.adsabs.harvard.edu/abs/2025MNRAS.537.1028P}
}

\newpage
\appendix
\renewcommand{\thesection}{\Alph{section}.\arabic{section}}
\setcounter{section}{0}
\normalsize

\end{document}